\documentclass[twocolumn,showpacs,a4]{revtex4}
\usepackage{graphicx}
\usepackage{bm}

 \def\barr{\begin{array}}
 \def\earr{\end{array}}
 \def\berr{\begin{eqnarray}}
 \def\err{\end{eqnarray}}
 \def\be{\begin{equation}}
 \def\ee{\end{equation}}
 \def\fr{\frac}
 \def\bs{\boldsymbol}

 \newcommand{\Schw}{Schwarzschild\ }
 \newcommand{\E}{Einstein\ }
 
 \renewcommand{\a}{\alpha}
 \renewcommand{\b}{\beta}

 \newcommand{\D}{\Delta}
 
 \renewcommand{\t}{\theta}

 \renewcommand{\v}{\varphi}
 
 \renewcommand{\O}{\Omega}

\begin{document}

 \title{Degeneracy in exotic gravitational lensing \\
 }

  \author{Margarita Safonova$^{a}$\thanks{rita@ducos.ernet.in}, and
 Diego F. Torres$^{b,c}$\thanks{dtorres@princeton.edu}}
 \address{
 $^a$Department of Physics and Astrophysics,
 University of Delhi, New Delhi--7, India}
 \address{ $^b$Physics
 Department, Princeton University, NJ 08544, USA}
 \address{ $^c$Lawrence Livermore National Laboratory,
7000 East Ave., Livermore, CA 94550, USA}

 \begin{abstract}
 We present three different theoretically foreseen, but unusual,
 astrophysical situations where the gravitational lens equation
 ends up being the same, thus producing a degeneracy problem. These
 situations are (a) the case of gravitational lensing by exotic
 stresses (matter violating the weak energy condition and thus
 having a negative mass, particular cases of wormholes solutions
 can be used as an example), (b) scalar field gravitational lensing
 (i.e. when considering the appearance of a scalar charge in the
 lensing scenario), and (c) gravitational lensing in closed
 universes (with antipodes).The reasons that lead to this
 degeneracy in the lens equations, the possibility of actually
 encountering it in the real universe, and eventually the ways to
 break it, are discussed.
 \end{abstract}
 \maketitle
 \section{Introduction}

 Gravitational lensing (GL) has long been advocated as an important
 tool in studying the Universe. It can act as a cosmic telescope,
 magnifying distant objects otherwise too dim to be detected. It is
 also an appropriate avenue to look for the detection of exotic
 objects in the universe. Unfortunately, initial hopes that
 GL would be able to resolve many long standing problems
 (e.g.~finding out the value of the Hubble constant through the
 time delay between the images, or giving an independent estimate
 of the masses of celestial bodies) went down as it was discovered
 that GL is subject to degeneracy and is highly model dependent.
 For example, when statistics of gravitational lensing was first
 introduced, it was hoped that the dependence of the image
 separations on the redshift of the source could constitute a test
 of the curvature of the universe \cite{statistics}. However,
 the large systematic and statistical uncertainties, involved
 with both the observed and predicted number of lensed arcs,
 as well as the rather small number of the multiply image systems available,
 do not allow us to constrain cosmological parameters based on current
 observations, nor even strongly favor one cosmological model above another
 \cite{waga,antipode,Cooray99}.

 An apparently unique feature of any lens model is its lens equation. Given a class
 of matter distribution, one can write a corresponding lens equation and
 solving it (when possible), find all necessary properties of a particular
 lens. In this work we present three different
 theoretically foreseen astrophysical situations possessing the
 same form of the lens equation. The degeneracy is presented among
 three cases: the exotic lens (understood as a lens made up of
 matter violating the weak energy condition), the lens endowed with
 scalar charge, and the case of a closed universe with a source
 behind the first antipode. We present the details of the resulting
 images configurations, the reasons leading to the degeneracy,
  and discuss some ways to get around it.

 \section{Cases}

 \subsection{Exotic lens}

 An exotic lens is a lens made up of matter violating the weak
 energy condition (WEC). Existence of such matter admits existence
 of negative energy densities---and so negative masses. Negative
 masses within General Relativity have been studied since Bondi's
 paper \cite{BONDI}, but recent attention was regained when
 wormhole solutions were presented \cite{motho,visser-book}.

 The detailed treatment of lensing properties of a point negative
 mass was presented in \cite{STR} (hereafter STR), for related
 studies see references therein. Here we will briefly describe the
 relevant features.  If $b$ is the impact parameter of the
 unperturbed light ray, the deflection angle for a negative point mass lens is
 ${\bs \alpha}_{\rm d} = -4 G |M| {\bf b}/c^2 b^2$ and the lens equation
 is~\footnote{According to the standard notations in the
 gravitational lensing, we define $D_{\rm L},\,D_{\rm S}$ and
 $D_{\rm LS}$ as the angular diameter distances to the lens, the
 source and between the lens and the source, respectively. $\beta$ is angular
 positions of the source and $\t$ is angular positions of the image.}
 \be
 \beta= \theta + \fr{4G|M|}{c^2}\fr{D_{\rm LS}} {D_{\rm S} D_{\rm
 L} \theta}\,\,.
 \ee
 Defining a useful angular scale in this problem, which in case of ordinary
 lensing is called an Einstein angle, \be \t_{\rm E}^2 = \fr{4 G|M|}{c^2}
 \fr{D_{\rm LS}} {D_{\rm S} D_{\rm L}}\,\,, \label{eq:einstein} \ee we rewrite
 the lens equation as $\beta=\theta+ \t_{\rm E}^2/ \t$. It can be solved
 to obtain two solutions for the image position $\theta$: $ \theta_{1,2} =
 \frac{1}{2} \left( \beta\pm \sqrt{\beta^2 - 4 \theta_{\rm E}^2}\right)$.
 Unlike in the lensing due to positive masses, three distinct regimes appear
 here: a) $\beta < 2 \theta_{\rm E}$.~There is no real solution for the lens
 equation---no images when the source is inside $2\theta_{\rm E}$; b) $\beta >
 2 \theta_{\rm E}$.~There are two solutions, corresponding to two images on
 the same side of the lens. One is always inside the angle $\theta_{\rm E}$,
 the other is always outside it; c) $\beta=2\theta_{\rm E}$.~This is a
 degenerate
 case, $\theta_{1,2}=\theta_{\rm E}$; two images merge at the $\theta_{\rm E}$
 angular radius, forming the {\it radial} arc.

 Two important scales in this case are the angle $\theta_{\rm
 E}$--- the angular radius of the radial critical curve (RCC), and
 the angle $2 \theta_{\rm E}$---the angular radius of the caustic. When the
 source crosses the caustic, the two images merge on the critical curve ($
 \theta_{\rm E}$) and disappear. We do not obtain a tangential critical curve
 (TCC)---in other words, no Einstein ring is possible and, accordingly, no
 tangential arcs.

 \subsection{Scalar fields}\label{sec:scalar}

 With increased interest in string theory, scalar fields, both
 minimally and conformally coupled to gravity, have been the
 subject of intensive research in recent years. Several possible
 astrophysical relevant features of scalar fields have been
 described. Among them the so-called `spontaneous scalarization'
 phenomenon
 \cite{damour-WHINNET}, the scalar field origin of dark matter on
 galactic (e.g. \cite{Guzman}) and cosmological scales (e.g.
 \cite{Cho}), quintessence (e.g. \cite{Fara}), the possible
 existence of supermassive scalar objects in centers of galaxies
 \cite{TORRES-GAL}, and, finally, the scalar field itself acting as
 a gravitational lens \cite{Vir,Matos,FRANZ}.

 One of the solutions to the Einstein-Massless Scalar field (EMS)
 equations derived for gravity minimally coupled to a scalar field
 is
 Janis-Newman-Winicour's (JNW) \cite{JANIS}. It describes the exterior of a static, spherically
 symmetric, and singular, massive object, endowed with the usual
 \Schw mass $M$ and a so-called scalar charge $q$---the signature
 of the conformal coupling of the massless scalar field with
 gravitation. The ``scalar charge" does not contribute to the total
 mass of the system, but it does affect the curvature of the
 spacetime \cite{vir97}.  The JNW solution also describes the space-time due to a naked
 singularity. The difference between these two objects lies in the
 value of a parameter $b$, which is defined as the curvature
 singularity ($b<r<\infty$, where $r$ is the radial coordinate of
 the JNW metric). When the radius of the object is greater than
 $b$, the lens is extended, spherically symmetric and static. When
 $r=b$, it is a naked singularity \cite{Vir}.

 The deflection angle (up to the second order) for the JNW metric is given by
 \begin{eqnarray} \a_{\rm d}(r_0) =
  \fr{4m}{r_0} + \fr{4 m^2}{r_0^2} \left( \fr{15\pi}{16}
 -2\right) +
  \fr{2}{r_0^2}\left[ 2m\sqrt{m^2+q^2} \right. \nonumber \\ \left. -\fr{q^2\pi}{8}\right] +
 \ldots\,\,, \end{eqnarray} with $m=GM/c^2$, $q$ the scalar charge
 and $r_0$ being the impact parameter.

 Another solution is an axially
 symmetric solution with a scalar field breaking the spherical symmetry, not
 the rotation \cite{Matos}. Authors find the deflection angle to be \be \a_{\rm d}(r_0)
 =\fr{4m}{r_0} + \fr{m^2}{r_0^2} \left( \fr{15 \pi}{4} -4 - \fr{1}{4}\pi
 R_{\rm sm}^2\right) + {\cal O}\left(\fr{m^3}{r_0^3}\right)\,\,, \end{equation}
 where the ratio of scalar charge to mass is denoted by $R_{\rm sm}$.

 We express the lens equations for both lenses in terms of $R_{\rm sm}$. For
 the JNW metric, the equation is \begin{eqnarray} \t -\b =\fr{D_{\rm LS}}{D_{S}}
 \left[\fr{4m}{D_{\rm L} \t} +   \fr{m^2}{D^2_{\rm L} \t^2} \left(\fr{15 \pi}{4}
 -8\right) + \right. \nonumber \\ \left. \fr{2m^2}{D^2_{\rm L} \t^2}\left\{2\sqrt{1+R_{\rm sm}^2} -
 \fr{\pi}{8} R^2_{\rm sm}\right\}\right] +\ldots \,\,. \end{eqnarray} And for the axial
 solution: \be \t -\b = \fr{D_{\rm LS}}{D_{S}} \left[\fr{4m}{D_{\rm L} \t} +
 \fr{m^2}{D^2_{\rm L} \t^2} \left\{\fr{15 \pi}{4} -4 -\fr{\pi R_{\rm
 sm}^2}{4}\right\}\right] + \ldots\,\,. \ee For very large values of $R_{\rm
 sm}$ \be R_{\rm sm} \equiv \fr{q}{M}\ge 500\,\,, \label{eq:ratio}
 \end{equation}
 in both cases the equations reduce to
 \be
 \b = \t -\fr{\t_{\rm E}^2}{\t}+ \fr{\t_{\rm E}^2}{\t^2} \fr{m\pi R_{\rm
 sm}^2} {16 D_{\rm L}  }\,\,,
 \label{eq:scalarleq}
 \ee
 where $\t_{\rm E}^2$ is the Einstein scale (Eq.~\ref{eq:einstein}).
 In both cases it was found that for the ``very large" ratio of the scalar
 charge to the mass (\ref{eq:ratio}) the lens forms two
 images of the opposite parities on the same side of the source. As
 $|\b|$ decreases, the two images meet at the RCC, forming the
 radial arc, and for any further decrease in $|\b|$ there are no
 images. There is no TCC (Einstein ring) for this case. For
 ``small" values of $R_{\rm sm}$ the lensing is qualitatively
 similar to the \Schw\/ lens. Even for smaller values of $R_{\rm
 sm}$ the qualitative behaviour of both scalar field lenses is
 similar; the quantitative differences appear only for small
 $R_{\rm sm}$ ($\simeq 15$) and for very small values of the image
 position $\b$ (a few milliarcseconds), rendering the possibility
 of distinguishing between different types of these lenses nearly
 impossible.

 \subsection{Closed Universe with Antipodes}\label{sec:antipode}

A recent renewed interest in closed universes
 followed the result that the most probable values of ($\Omega_{\rm
 M}$, $\Omega_{\Lambda}$) from observations of high redshift
 supernovae are indeed consistent with a mildly closed Universe
 (\cite{perm,turok}), though the extreme closed model having
 antipodal redshift of $z_{\rm antipode} < 4.92$ is ruled out
 \cite{antipode}.


 The metric for the closed FRW universe is given by $ ds^2 =
 c^2dt^2-a^2(t)\left(d\chi^2 +\sin^2{\chi}\left( \sin^2{\t}d\v^2 +
 d\t^2\right) \right)$, where $\chi$ is the conformal radial
 coordinate, which takes values in the interval $0 \le \chi \le
 2\pi$ and is related to the comoving radial coordinate $r$ by $r=\sin{\chi}$.
 If $D_{\rm A}(z)$ is the angular diameter distance to an object at redshift
 $z$, then \be D_{\rm A}(z)=\fr{cH_0^{-1}}{(1+z)\left|\O_{\rm total} -1
 \right|^{\fr{1}{2}}} \sin{\chi}\,\,, \label{eq:D_A} \ee \be
 \chi(z)=\left|\O_{\rm total} -1 \right|^{\fr{1}{2}}\int_0^z\fr{du}{H(u)/H_0.
}\,\,.
 \ee Angular distance $D_{\rm A}(z)$ becomes zero at the points
 $\chi(z_n^a)=(2n-1)\pi/2$, where $n=1,2,\ldots$. These points are called the
 antipodal points and the corresponding redshifts, $z_n^a$,
 antipodal redshifts. The effect of the closed geometry is to focus the light
 from any object in the opposite hemisphere.


 Lensing in a universe with antipodes was first described in \cite{gott},
 and later, by Saini \cite{saini}. Consider the situation of a source beyond
 the first antipode and the lens much closer than it, $0 <\chi_{lens} <\pi<
 \chi_{source}$. In this case $D_{\rm S}$ is negative (see Eq.~\ref{eq:D_A}),
 while $D_{\rm L}$ and $D_{\rm LS}$ could still be positive. This makes
 $\Sigma_{\rm crit}$ negative and, hence, the convergence, $\kappa
 =\Sigma({\bs\theta})/\Sigma_{\rm crit} <0$. We remind that
 $\Sigma(\bs{\theta})$ is the two-dimensional surface mass density
 of the lens and $ \Sigma_{\rm crit} = ({c^2}/{4\pi G}) ({D_{\rm
 S}}/{D_{\rm L} D_{\rm LS}})$ is the critical density. A point lens
 will still form two images of the background sources, though they
 would be on the same side of the lens, unlike the case of normal
 lensing when they straddle the lens on either side. The lens
 equation for this case is \be \beta = \theta +
 \frac{\theta_E^2}{\theta}\,\,, \label{eq:leq} \ee with the same
 angular scale as before (Eq.~\ref{eq:einstein}), and the solutions
 are, again, $ \theta = \frac{1}{2}\left[\beta \pm \sqrt{(\beta^2-4
 \theta_E^2}) \right]. $ From this it is clear that no images are
 formed if $\beta < 2\theta_E$, and outside this radius both the
 images are on the same side of the lens. No Einstein ring can be
 formed: for $\b=0$ there is no solution.

 \section{Discussion}


 In the three considered cases the reasons for the same observable signatures
 are different. In the first case, it is due to the assumption of the negative
 sign of the mass term and the corresponding negative sign
 of the deflection angle. In the second case, the scalar charge has an
 effective `negative' contribution to the
 space-time
 curvature around the object. In that way, it is doing the same job
 as the negative mass, deflecting the light from it.
 Finally, in the third case, it is due to the negative sign of the
 angular diameter distance, leading to the negative sign of the
 convergence $\kappa$. To be more precise, the basic form of lens
 equation is $\bs{\b} =\bs{\t} - \bs{\nabla} \psi(\bs{\t})$, where the
 dimensionless relativistic lens potential $\psi$ satisfies the
 two-dimensional Poisson equation $\D \psi(\bs{\t}) = 2 \kappa (\bs{\t})$ (see,
 eg.
 \cite{schneider}). Thus, negative $\kappa$ leads to a lens equation of the
 form (\ref{eq:leq}). The question now is how to differentiate among the
 different possibilities if we actually see the effects of (\ref{eq:leq}).

 If we measure the redshift of the images, this could
 help getting rid of the antipodes case. The fact that a `normal'
 multiple imaged quasar exists at a redshift of $z=4.92$ indicates
 that if the putative system is at a lower redshift, the case for
 the closed universe is ruled out (though, of course, not ruling
 out that we might, in fact, live in a closed universe).

\begin{figure}
\includegraphics[width=0.4\textwidth,height=9cm]{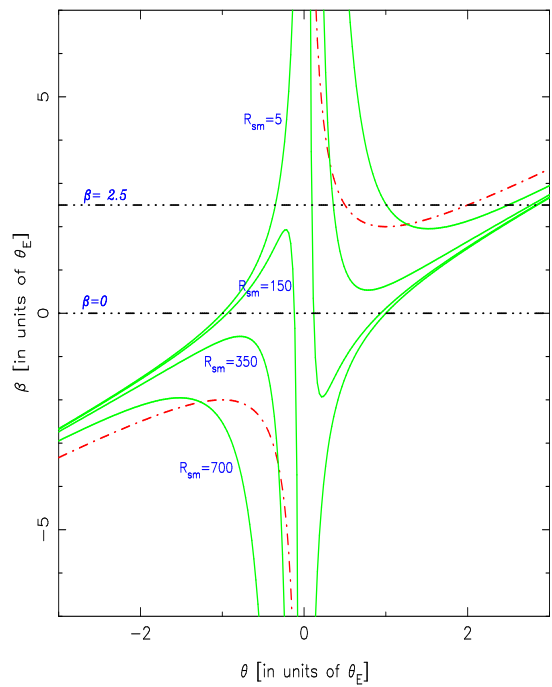}
\caption{Lensing equation curves for the lens with the scalar
 charge (solid lines) for four values of $R_{\rm sm}$ and the
 negative mass lens (dashed-dotted lines). Horizontal
 dot-dot-dashed lines represent a constant source position. Here $|M|_{\rm
 lens} =10^{11}\,M_{\odot}$, $z_{\rm source}=3.0$, $z_{\rm lens}=1.3$.}
 \label{fig:leq}
\end{figure}

 In Fig.~\ref{fig:leq} we compare the lensing equation curves for
 the scalar field lens with four different values of scalar charge
 (solid lines) and a negative mass lens (dashed-dotted line). The
 lens equation curves are symmetric with respect to the origin; we
 label only the left part of the scalar field lens curves. The
 horizontal dot-dot-dashed lines are the lines of a constant source position
 and their intersection with the lensing curves shows the positions and
 numbers of images. For a small value of a scalar charge ($R_{\rm sm}=5$) the
 lens
 behaves like a \Schw lens. Though the forms of the lens equations for these
 two
 cases are different, we can see from the figure that with
 increased value of a scalar charge ($R_{\rm sm} \ge 300$), the
 curves become similar (the positive term in Eq.~\ref{eq:scalarleq} begins to
 dominate). The situation with intermediate $R_{\rm sm}$ values (and formation
 of two \E rings) is very unstable and with a slight change of lens system
 parameters (e.g. mass of the lens) quickly relaxes in one of the
 two other cases. Thus, in order to mimic the effect of an exotic lens, the
 scalar charge must be large. However, it is not at all clear whether
 the value of the scalar charge should be large or small.

 For the case of negative mass objects, the lens will be, most
 likely, not seen. We would rather expect the existence of compact
 objects of solar or sub-solar negative mass, as opposed to larger
 structures whose effects are apparently absent from all deep
 fields images \cite{macro}. Then we will not be able to see
 the images, but only to observe the microlensing light curves, and/or
 distinctive chromaticity effects (see \cite{eiroa}).

 Finally, not only gravitational lensing observations can break
 this degeneracy. Forthcoming supernovae studies,
 might clearly distinguish between the different
 FRW models. If we live
 in a flat or open FRW universe, one of the three cases herein
 treated disappears. Theoretical investigations in GR and related
 areas might eventually lead to the proof of the cosmic censorship conjecture,
 or the
 energy conditions, and in that case, neither the JNW solution with non-zero
 scalar charge nor exotic negative
 mass lenses will be realized. As for the time being, we
 can not neglect any of these theoretical situations and the best
 way to attack the problem would be imposing direct astrophysical
 bounds. The MOA group is currently adapting their alert systems
 to take into account this kind of exotic lensing \cite{P}.
 Maybe great surprises await to be discovered.

A number of dark lens candidates (quasar
 pairs with no detectable lensing mass) have been reported in the
 literature, and their true nature is still a matter of an open
 vigorous debate \cite{dark}. Current attempts at solving the problem by
 trying to fit the existing candidates under the lensing by empty dark matter
 halos do not work \cite{rusin02}. Dark lenses are expected to have very
 small magnification ratios and prominent third images. The majority of
 current candidates have flux ratio that differ significantly from unity,
 and do not feature any third image. The absence of quads
 (four-image systems) makes the current dark lens sample even more
 peculiar. The type of lenses described here, with their image
 configurations (always two images, albeit on one side of most
 probably invisible lens), and magnification ratios (the ratio is
 much steeper than for the equivalent ordinary lensing, seem to be an interesting
 possibility for the dark lenses.

  The considered cases might be not the only
diverging lens systems possible. According to the paper by
Amendola et al. \cite{waga2}, huge empty voids with radii larger
than 100 $h^{-1}$ Mpc can be individually detected via diverging
weak lensing. Empty voids with radii 30 $h^{-1}$ Mpc,
characteristic of those seen in galaxy redshift surveys, have a
lensing signal to noise ratio smaller than unity.  Finally, we
note that we have considered only strong lensing events, because
it is only in this case that the degeneracy is manifested.



 \acknowledgments MS is supported by a ICCR scholarship
 (Indo-Russian Exchange programme) and wishes to thank Dr. Amber Habib for
 his mathematical insights.

 \end{document}